\documentclass[a4paper,11pt]{article}
\usepackage[utf8]{inputenc}
\usepackage{authblk}
\usepackage{subfig}
\usepackage{jinst}
\usepackage{ifthen}
\usepackage{booktabs}
\usepackage{lineno}
\newboolean{inbibliography}
\setboolean{inbibliography}{false}
\newboolean{pdflatex}
\setboolean{pdflatex}{true}
\newboolean{articletitles}
\setboolean{articletitles}{true}
\usepackage{soul}
\usepackage{natbib}
\usepackage{graphicx}
\usepackage{dirtree}
\usepackage{siunitx}
\usepackage{amssymb}
\usepackage{amsmath}
\usepackage{hyperref}
\usepackage[percent]{overpic}
\usepackage[table]{xcolor}
\usepackage{mciteplus}

\title{Fast simulation of muons produced at the SHiP experiment using Generative Adversarial Networks}

\collaboration{
SHiP collaboration
 \begin{flushright}
  \includegraphics[height=20mm]{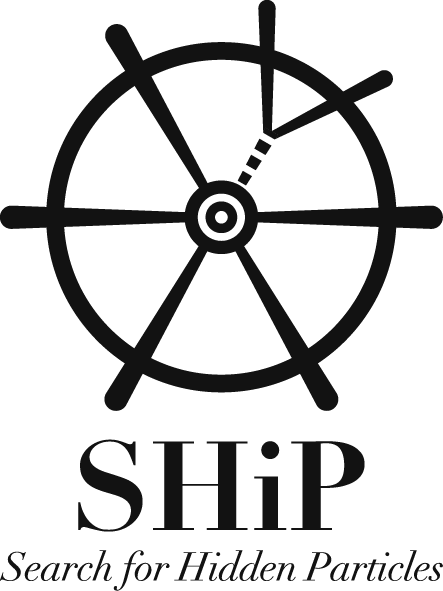}
 \end{flushright}
}

\author{
\begin{flushleft}
C.~Ahdida$^{44}$,
A.~Akmete$^{48}$,
R.~Albanese$^{14,d,h}$,
A.~Alexandrov$^{14}$,
A.~Anokhina$^{39}$,
S.~Aoki$^{18}$,
G.~Arduini$^{44}$,
E.~Atkin$^{38}$,
N.~Azorskiy$^{29}$,
J.J.~Back$^{54}$
% ...

\vspace*{0.5cm}

{\footnotesize \it

% ...
$ ^{14}$Sezione INFN di Napoli, Napoli, Italy\\
% ...
$ ^{18}$Kobe University, Kobe, Japan\\
% ...
$ ^{29}$Joint Institute for Nuclear Research (JINR), Dubna, Russia\\
% ...
$ ^{38}$National Research Nuclear University (MEPhI), Moscow, Russia\\
$ ^{39}$Skobeltsyn Institute of Nuclear Physics of Moscow State University (SINP MSU), Moscow, Russia\\
% ...
$ ^{44}$European Organization for Nuclear Research (CERN), Geneva, Switzerland\\
% ...
$ ^{48}$Middle East Technical University (METU), Ankara, Turkey\\
% ...
$ ^{54}$University of Warwick, Warwick, United Kingdom\\
% ...
$ ^{d}$Universit\`{a} di Napoli ``Federico II'', Napoli, Italy\\
% ...
$ ^{h}$Consorzio CREATE, Napoli, Italy\\
% ...
}
\end{flushleft}
}

\abstract{
     This paper presents a fast approach to simulating muons produced in interactions of the SPS proton beams with the target of the SHiP experiment. The SHiP experiment will be able to search for new long-lived particles produced in a 400~GeV$/c$ SPS proton beam dump and which travel distances between fifty metres and tens of kilometers. The SHiP detector needs to operate under ultra-low background conditions and requires large simulated samples of muon induced background processes.  Through the use of Generative Adversarial Networks it is possible to emulate the simulation of the interaction of 400~GeV$/c$ proton beams with the SHiP target, an otherwise computationally intensive process. For the simulation requirements of the SHiP experiment, generative networks are capable of approximating the full simulation of the dense fixed target, offering a speed increase by a factor of  $\mathcal{O}(10^6)$. To evaluate the performance of such an approach, comparisons of the distributions of reconstructed muon momenta in SHiP's spectrometer between samples using the full simulation and samples produced through generative models are presented. The methods discussed in this paper can be generalised and applied to modelling any non-discrete multi-dimensional distribution.
}
\begin{document}

\arxivnumber{1909.04451}
\emailAdd{alex.marshall@cern.ch}

\maketitle

\section{Introduction}\label{intro}
    
    Generative networks are a class of machine learning algorithms designed to generate samples according to a multidimensional function, given a randomly distributed input sample. Generative networks have been studied in the machine learning community primarily for the purpose of image generation. Each image in a training set is made up of a multitude of pixels, corresponding to a data point in a high dimensional space. Within this space, underlying features of the set of images are encoded through dependencies between pixels. Generative networks attempt to model the characteristics that define a specific set of training images. These models can then be used to generate images that are faithful emulations of the original training set. Generative networks have been successfully employed for a variety of applications such as: generating high quality images that obey fundamental features of training set images; the generation of images from descriptive text~\cite{zhang2017stackgan}; modelling image captions~\cite{pu2016variational}; producing photo realistic super resolution images~\cite{ledig2017photo}; and generating high resolution images from semantic mapping~\cite{wang2018high,isola2017image}. 

    Searches for physics beyond the Standard Model often involve looking for rare signatures and must therefore be able to suppress background processes which can be many orders of magnitude more abundant than the signal. In order to optimise the design of the detectors, develop reconstruction algorithms and understand the efficiency of the selection criteria, large samples of simulated background events are required. Dedicated software packages such as \texttt{GEANT4}~\cite{ref:geant4} model the transport of particles through the material and the detector response. In many cases, the CPU requirements to simulate these interactions with matter prohibit the production of large numbers of background events due to the computationally expensive procedure. Therefore, the computing demands of the simulation of high energy physics experiments are increasing exponentially~\cite{albrecht2018roadmap}. Recent algorithmic improvements that take advantage of high performance computing resources aim at reducing simulation time, resulting in an order of magnitude increase in speed~\cite{Canal:2016dki}. This improvement is not sufficient to meet the demands of future particle physics experiments, such as those at the High Luminosity LHC, for large simulation samples~\cite{albrecht2018roadmap}. Generative neural networks offer an alternative approach to simulation by modelling non-analytical functions in a computationally efficient way~\cite{mirza2014conditional}.
    
    The use of generative networks for particle physics originally focused on image based generation. Examples of their application include: the development of Location Aware Generative Adversarial Networks for the production of images of jets~\cite{de2017learning}; the simulation of  reconstructed cosmic ray induced air showers~\cite{erdmann2018deep}, and of showers in electromagnetic calorimeters~\cite{caloGAN}; the fast simulation of Cherenkov detectors~\cite{derkach2019cherenkov}. More recently, generative approaches have also been adopted to simulate the kinematics of final state particles emerging from physical exclusive two-to-two processes, such as $Z$ or top-quark production at the LHC~\cite{LHCGAN,Otten:2019hhl}. In Ref.~\cite{Hashemi:2019fkn}, generative networks were used to simulate the detector reconstruction of the $Z\to\mu^+\mu^-$ process at the LHC. 
    
    This paper describes the use of generative networks to emulate the kinematics of muons produced through the interactions of high energy protons with the dense target designed for the Search for Hidden Particles (SHiP) experiment~\cite{ship1504technical}. This approach offers a gain of multiple orders of magnitude in the computational efficiency of such processes. In contrast to the aforementioned use of generative networks to approximate a single exclusive process, this work employs four different generative networks to model the kinematics of muons originating from a multitude of processes, including muons from secondary interactions or particle showers in SHiP's target. Large samples of generative based muons can then be passed through the \texttt{GEANT4} based simulation of the rest of the SHiP experiment, offering a precise modelling of its detector response and reconstruction. 

    This paper is organised as follows: Sections~\ref{sec:shipdetector} and~\ref{sec:ship} describe the SHiP detector and its simulation framework respectively; Section~\ref{sec:genmodels} discusses the generative models used in this analysis, while Section~\ref{sec:training} details how these networks are trained and optimised for the SHiP experiment. Section~\ref{sec:generatedmuons} and \ref{sec:comp} then present the performance of the generative models in simulating muons produced through interactions of high energy protons with the SHiP target, compared to the \texttt{Pythia8}~\cite{ref:Pythia8} and \texttt{GEANT4} frameworks. Finally, Section~\ref{sec:bench} discusses the computational time required to produce muons through generative networks.
    
\section{The SHiP experiment}\label{sec:shipdetector}

\begin{figure}[t!]
        \centering
        \begin{overpic}[width=0.85\textwidth,keepaspectratio]{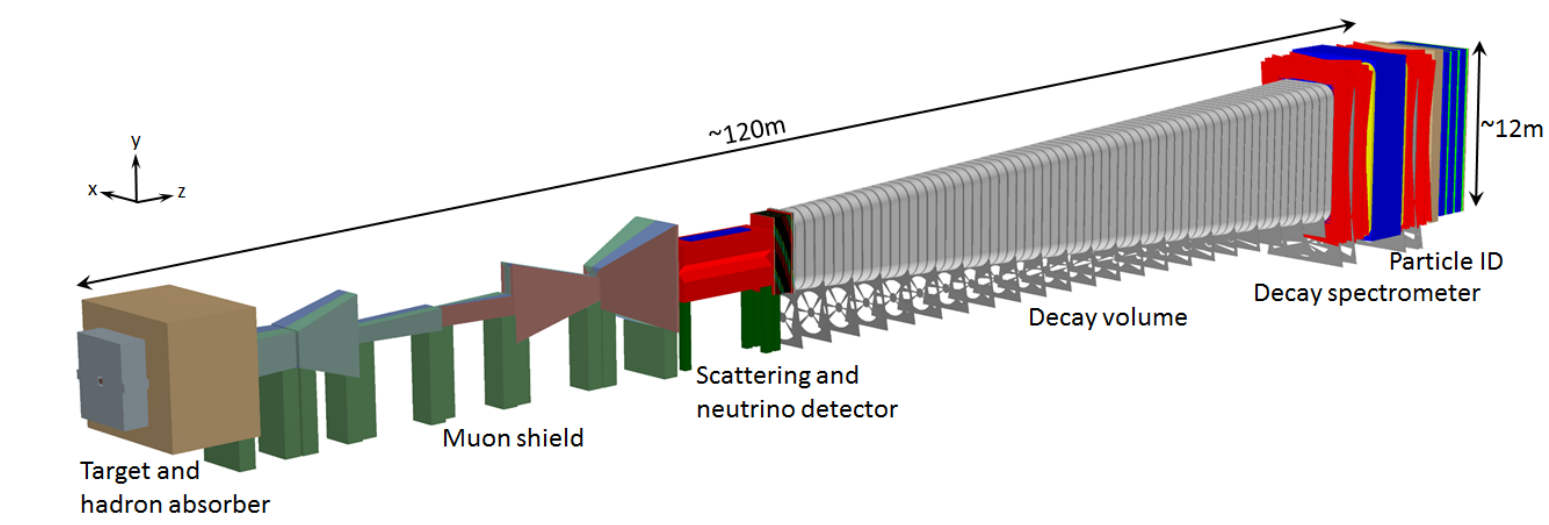}
        \put(83,0){\includegraphics[width=0.075\textwidth,keepaspectratio]{SHiP-Full_Black.png}}  
        \end{overpic}
        \caption{SHiP facility layout \cite{ship2019ship}.}
        \label{fig:SHiP}
    \end{figure}

   The Search for Hidden Particles experiment (SHiP) is a proposed experiment that will operate at the the prospective general purpose fixed target facility at the CERN Super Proton Synchrotron (SPS) accelerator. The SHiP experiment aims to search for long-lived exotic particles with masses between a few hundred MeV$/c^2$ and a few GeV$/c^2$. These particles are expected to be produced in the decays of heavy hadrons. The facility is therefore designed to maximise the production rate and detection efficiency for charm and beauty mesons and their decay products, while maintaining the lowest possible background rate. The 400~GeV$/c$ proton beam extracted from the SPS will be dumped on a high density W/Mo target with the aim of accumulating $2\times 10^{20}$ protons on target during 5 years of operation. The charm production at SHiP will exceed that of any existing or planned facility. The SHiP detector, shown in Fig.~\ref{fig:SHiP}, incorporates two complementary apparatuses, the Scattering and Neutrino Detector (SND), and the Decay Spectrometer (DS). The SND will be used to search light dark matter particles, and perform neutrino physics measurements. The DS aims at measuring the visible decays of hidden sector particles by reconstructing their decay vertices in a 50~m long decay volume, making use of a magnetic spectrometer, veto systems and particle identification detectors. Further details of the design of the detector can be found in Ref.~\cite{ship2019ship}. Such a setup will allow the SHiP experiment to probe a variety of models that predict light long-lived exotic particles. 
    
    Since particles originating in charm and beauty meson decays are produced with a significant transverse momentum with respect to the beam axis\footnote{In the SHiP coordinate system the z-axis is along the beam line and the y-axis is pointing upward.}, the detector is placed as close as possible to the target. The high flux of muons produced in the target represents a serious background in searches for hidden particles. A critical component of the SHiP experiment is the muon shield~\cite{akmete2017active}, which deflects muons produced in the target away from the detector placed downstream of the target. The SHiP detector is designed to reconstruct the exclusive decays of hidden particles and to reduce the background to less than 0.1 events in the full five year period of operation. 

\section{The SHiP simulation}\label{sec:ship}

    The simulation of the various physics processes of the response of the SHiP detector are handled by the \texttt{FairShip} software suite, which is based on the \texttt{FairRoot} software framework~\cite{al2012fairroot}. Within \texttt{FairShip}, primary collisions of protons are generated with \texttt{Pythia8} and the subsequent propagation and interactions of particles are simulated with \texttt{GEANT4}. Neutrino interactions are simulated with \texttt{GENIE}~\cite{ref:GENIE}, while heavy flavour production and inelastic muon interactions with \texttt{Pythia6}~\cite{ref:Pythia6} and \texttt{GEANT4}.  Secondary heavy flavour production in cascade interactions of hadrons originating from the initial proton collision with the SHiP target is also taken into account~\cite{ref:cascade}. The pattern recognition algorithms used to reconstruct tracks from the hits on the strawtubes of the DS are described in~\cite{ref:ship_spectrometer}, and the algorithms for particle identification are presented in~\cite{ref:ship_pid}.
    
    In order to optimise the design of the active muon shield, and develop the reconstruction and selection algorithms of the SHiP experiment, a large simulation campaign was undertaken. Muons produced in the SHiP target were simulated with momentum $p>10$~GeV$/c$ and a sample corresponding to approximately $10^{11}$ protons on target was produced. In order to enhance this sample with muons likely to enter the DS, the cross-section of muons produced from decays of $\rho^0$, $\omega$, $\eta$ and $\eta^{\prime}$ mesons was enhanced by a factor of 100. Similarly, the cross-section for photon conversions into muon pairs was also enhanced by the same factor. The full simulation of this sample, corresponding to a fraction of the $4\times10^{13}$ protons-on-target SPS spill, required months of running on dedicated CPU farms. An order of magnitude increase of this sample could be achieved by exploiting symmetries of the system, such as that in the azimuthal plane of the collision. However, generating even larger samples commensurate to the $2\times 10^{20}$ SPS protons on target expected during the lifetime of the experiment is impossible using conventional particle simulation methods. The simulation of the initial proton interaction with the SHiP target, including the subsequent secondary interactions of particles with the target and the hadron absorber, requires significant computing power. Methods such as \texttt{SMOTE}~\cite{chawla2002smote} and \texttt{ADASYN}~\cite{he2008adasyn} could be used to synthesise a sample of muons that is larger than the original fully simulated sample. These methods rely on producing muons whose position and momentum vectors take values that lie in between those of existing muons in the fully simulated sample. Generative adversarial networks can offer an alternative way of producing orders of magnitude larger samples with minimal expense to the fidelity of the generated muons. 

\section{Generative adversarial networks}\label{sec:genmodels}

    Neural networks model functions that map an $n$-dimensional input parameter space into an $m$-dimensional output, and are widely employed in the particle physics community. A traditional neural network is built up of multiple \textit{layers}: an input layer, one or more intermediate hidden layers, and an output layer. Layers are built from many individual \textit{nodes}, and a pattern of connections joins nodes in adjacent layers. Each node has an associated tunable \textit{bias} term that acts as an activation threshold of the node, and each connection has an associated tunable \textit{weight} representing the strength of the connection. The simplest pattern of connections between layers is one were the nodes in one layer are fully connected with nodes in the adjacent layer. In this configuration the output value of each node is calculated by firstly calculating the sum of the output values of each node from the previous layer, weighted by the strength of each connection. This weighted sum is then shifted by the bias term and passed through an activation function that modulates the output of a node. Depending on the layer that a particular node belongs to, different types of activation functions can be used. For instance, hidden layers often make use of the so called ``leaky rectified linear unit" function~\cite{maas2013rectifier} and in the final layer a sigmoid function could be used to transform the output into a value between 0 and 1. This choice would be appropriate in a binary classification network, whose output is an estimate of whether the input sample originated from one out of two classes of samples.
    
    A neural network must be trained in order for it to successfully approximate a function. The training process involves tuning and updating the weight and bias parameters of the network, with ``supervised learning" being the most traditional approach to training. In the first stage of a binary classification problem, labelled data are passed through the network. Output values are then recorded and compared to the true labels through the use of a \textit{loss function}. The loss function provides a quantitative measure of the network's performance on a set of input training samples. A large value of the loss function indicates that the network is unable to distinguish between the two classes of samples. The value of the loss function is used in a process called \textit{back-propagation} to update the weight and bias parameters across the network in an effort to improve the network's performance~\cite{rumelhart1988learning}. Neural networks are trained in steps, where in each step a small \textit{batch} of training data is used, the loss function is then evaluated using this batch of data, and the weight and bias parameters of the network are updated for the next step. 
    
    Generative Adversarial Networks (GANs) employ two competing neural networks, one acting as a generator and the other as a discriminator~\cite{goodfellow2014generative}. The generator $\hat{G}$ is trained to map an input vector of random noise $z$ to an output generated vector $G(z;\theta_g)$, where $\theta_g$ are parameters of the network and the dimensionality of $z$ is typically larger than that of $G(z;\theta_g)$. The discriminator $\hat{D}$, with trainable parameters $\theta_d$, is trained to map an input vector $x$ to an output prediction $D(x;\theta_d)$, which is a value between 0 and 1. In the study presented in this paper $G(z;\theta_g)$ and $x$ represent the momentum and position vector of the muons. The value of $D(x;\theta_d)$ represents the probability that $x$ originated from the training sample. A value of $D(x;\theta_d)$ closer to 0 indicates that $\hat{D}$ expects the sample to have been generated by $\hat{G}$, whereas if $D(x;\theta_d)$ is closer to 1 $\hat{D}$ is predicting that the sample originated from the training data. 
    
    The discriminator and generator networks are trained using an iterative approach. Firstly, the discriminator is trained to distinguish between generated and training samples via a binary crossentropy loss function $L_d$. This is a common loss function for training classifier networks and is defined as
        \begin{equation}
        L_d = -[y_{\mathrm{true}} \log({y_{\mathrm{pred}}}) + ( 1 - y_{\mathrm{true}} ) \log( 1 - y_{\mathrm{pred}})],
        \end{equation}
    where $y_{\mathrm{true}}$ takes the values of 1 or 0 for the training or generated label of the sample respectively, and $y_{\mathrm{pred}}$ is the predicted label by the discriminator given by $y_{\mathrm{pred}}~=~D(x;\theta_d)$. The value of this loss function increases rapidly the further $y_{\mathrm{pred}}$ is from the $y_{\mathrm{true}}$. Large values of the loss function bring significant changes in the values of trainable parameters $\theta_d$ in the network.

    The generator network is then trained in a stacked model which directly connects the output $x_{\mathrm{gen}}$ of $\hat{G}$ to the discriminator prediction $D(x_{\mathrm{gen}};\theta_d)$. This is the adversarial component of the GAN, it is only the feedback of $\hat{D}$ that influences the training of $\hat{G}$. The $x_{\mathrm{gen}}$ never directly affects the training of $\hat{G}$. In this stacked model all training parameters of the discriminator, $\theta_d$, are fixed to the values obtained from the previous training step of $\hat{D}$. The trainable parameters, $\theta_g$, of the generator are updated based on the loss function, $L_g$, whose value depends on the output of the discriminator and is defined as
        \begin{equation}\label{L_g}
        L_g = -\log(D(x_{\mathrm{gen}};\theta_d)).
        \end{equation}
    Low values of $D(x_{\mathrm{gen}};\theta_d)$ indicate that the discriminator is confident that the sample $x_{gen}$ originated from the generator, leading to a large value of $L_g$. Generated samples that closely resemble training samples will return higher values of $D(x_{\mathrm{gen}};\theta_d)$ and consequently lower values of $L_g$ as the discriminator is successfully tricked into guessing a sample originated from the training sample. 
     
     The training of the discriminator and generator is repeated using the samples generated by the previous step. The training of the GAN is completed when generated samples $G(z;\theta_g)$ are indistinguishable from training samples. Different approaches are employed to determine the end of the training. In this paper the metric used to monitor the quality of the training is discussed in Sec.~\ref{fom}. The overall accuracy of the generator depends on how well the discriminator is trained to distinguish between generated and training samples.

\section{GANs for the SHiP experiment}\label{sec:training}

    The GAN is trained on a sample of \num{3.4e8} muons passing through the target and hadron absorber in the full simulation campaign discussed in Sec.~\ref{sec:ship}. As mentioned, this training sample is artificially enriched with muons from rare processes. Therefore, in order to obtain a physical admixture of muons from various sources, batches of muons are extracted from the training sample according to a probability that corrects for this enhancement.

    Training is performed on the position $\textbf{r}$ and momentum $\textbf{p}$ vectors of these muons at their point of production. Therefore, the GAN generates position and momentum vectors of muons at their production point within the target. Subsequently, they are propagated through the active muon shield and the Decay Spectrometer, relying on \texttt{GEANT4} to simulate muon interactions with matter. This procedure allows for a fast production of large muon samples, while maintaining the flexibility to optimise the magnetic shield and downstream detector elements of SHiP, as well as the ability to correct for effects due to the spatial distribution of the proton beam impinging on the target. 
    
    Four separate GANs are trained, separated by muon charge and prompt or displaced origin. The $x$- and $y$-coordinates of muons originating from prompt decays of mesons such as the $\rho^0$, $\phi$ and $J/\psi$ are always the same. This is a consequence of the training sample that relies on \texttt{Pythia} with no smearing of the proton-beam distribution. As such, muons from prompt sources are treated separately from muons originating from other sources. Therefore, the GANs trained on prompt muons generate four features ($z$, $\textbf{p}$), and the GANs for non-prompt muons generate six ($\textbf{r}$, $\textbf{p}$). In this approach correlations between muons produced in pairs from, for example, vector-meson decays are ignored. Muons are generated individually and any correlation is assumed to be lost via the multiple scattering of the muons through the hadron absorber and muon shield. 
    
    \subsection{Pre-processing}\label{pre_process}
        
        The distribution of the $x$- and $y$- coordinates of muons from non-prompt sources is extremely peaked around the interaction point. Therefore, each value of the $x$ ($y$) distribution $x^i$ ($y^i$) is transformed as
            \begin{equation}
            x^i_{\mathrm{trans}}=\begin{cases}
            -\sqrt{|x^i - \overline{x}|} & \text{if $x^i<0$},\\
            \sqrt{|x^i - \overline{x}|} & \text{if $x^i>0$},
            \end{cases}
            \end{equation}
        before training the GANs. This transformation widens the distributions, which proves easier for the GANs to model. The distributions of all the input features are then normalised to values between -1 and 1. This transformation is reversed to obtain physical values of the generated output.

    \subsection{Figure of merit}\label{fom}
    
        An important requirement of the full simulation of the SHiP detector is to accurately model the flux of muons reaching the Decay Spectrometer. This flux crucially depends on the momentum distribution of the muons entering the muon shield of the SHiP experiment. Therefore, muons generated through the GAN approach must closely match the kinematic distributions of the muons produced in the target using the full simulation. 
        
        In order to optimise the architecture of the networks and to quantify the quality of the training procedure a figure of merit, FoM, is developed with the following requirements. The FoM must account for how well the GAN is able to model individual features and the correlations between them. Furthermore, it is important that the FoM offers an independent metric of the quality of the training of the GAN since the discriminator and the generator of the GAN improve in tandem during the iterative training procedure. Finally, the calculation of the FoM must be fast so that it does not slow down the training process. 
        
        During the training process, small test samples are generated to test the progress of the procedure. As the number of muon features can span a six dimensional space, a small generated sample of muons results in a sparsely populated feature space. Therefore,  traditional binned goodness of fit methods, such as $\chi^2$-tests, break down as almost all bins in this space have a low occupancy. Boosted decision trees can overcome this issue ~\cite{weisser2016machine} and satisfy the aforementioned requirements on the FoM. 
        
        A gradient boosted decision tree, BDT, is trained periodically to distinguish between generated and fully simulated muon samples. The BDT uses 100,000 muons generated from the latest GAN configuration and 100,000 randomly selected, fully simulated muons. Half of the muons in each sample are used for training and the other half for testing. The resulting performance of the BDT is quantified through the area under the receiver operating characteristic curve (ROC AUC). A generated sample that is indistinguishable from a fully simulated sample would return an ROC AUC value of 0.5.

    \subsection{Network optimisation and GAN architecture}\label{sec:networkop}
        
        All networks are trained with a mini-batch gradient descent approach~\cite{ruder2016overview} and at each training step the networks use a sub-sample of training data. The generators and discriminators of the GANs are built using only fully connected layers, resulting in a GAN performance that is independent of the ordering of the muon parameters in the vectors of features that make up the training sample. The number of nodes, the size of the batch sub-sample, the number of layers and the learning rate of the networks are coarsely optimised through a grid search over these parameters. The four GANs are trained until the ROC AUC of the BDT based FoM described in Sec.~\ref{fom} flattens out, and the selected architecture is that which minimises the ROC AUC of the FoM.

        \begin{figure}[t!]
            \centering
            \includegraphics[width=\textwidth,height=\textheight,keepaspectratio]{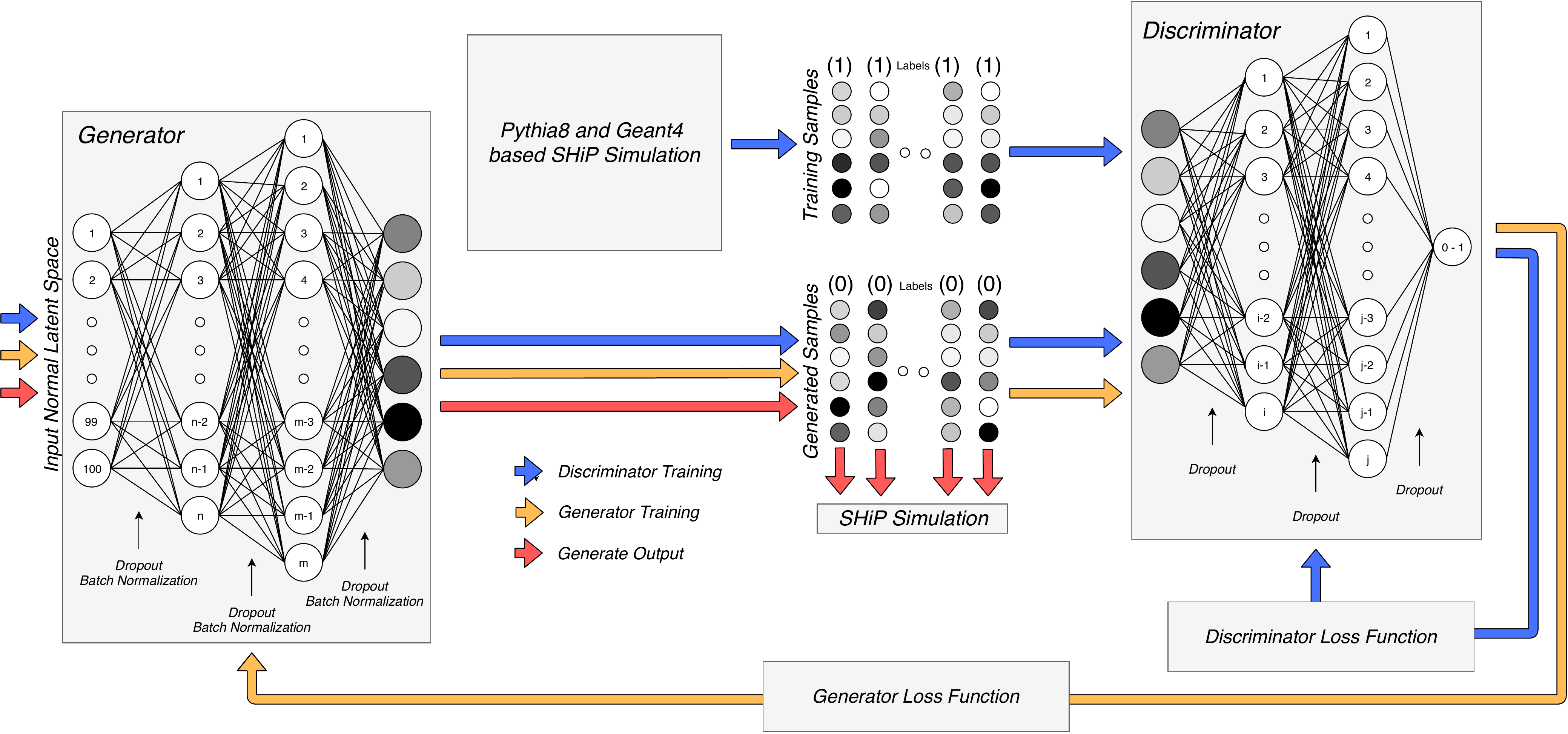}
            \caption{Optimal GAN architecture obtained for the simulation of muon background in SHiP. The number of nodes in each layer for prompt $\mu$ and non-prompt $\mu$ GANs are given in the text. Arrows indicate the flow of samples and loss information for each stage of training and generation. The features in the generated and training samples can take values between -1 and 1 as denoted by the varying shades of grey.
            }
            \label{GAN-diagram}
        \end{figure}
            
        As a result of this optimisation procedure, the GANs for both prompt and non-prompt muons follow the architecture shown in Fig.~\ref{GAN-diagram}. Leaky rectified linear unit activation functions are used at every hidden layer.  The generator and the discriminator have two hidden layers in an inverted pyramidal structure. For the prompt muon GANs, the number of nodes in each hidden layer of $\hat{G}$ are 1536 and 3072 and for $\hat{D}$ are 768 and 1536. For the non-prompt GANs, the number of corresponding nodes are 512 and 1024 for $\hat{G}$ and 768 and 1536 for $\hat{D}$. The input to the generators relies on sampling from a latent space given by a 100 dimensional unit Gaussian distribution. The last layer of $\hat{G}$ has a $\tanh$ activation function in accordance to the transformed range of the input features described in Sec.~\ref{pre_process}. The last layer of $\hat{D}$ has a sigmoid activation function providing an output between 0 and 1 that represents $\hat{D}$'s judgement on the origin of a sample. Dropout layers with a dropout probability of 0.25 are added between each layer of $\hat{G}$ and $\hat{D}$ to prevent overfitting~\cite{srivastava2014dropout}. Batch normalisation layers are also added between layers of $\hat{G}$~\cite{ioffe2015batch}.
        
        For this study the \texttt{Adam} optimisation algorithm~\cite{kingma2014adam} was used in training the networks. Employing the \texttt{AMSgrad} algorithm with the \texttt{Adam} optimiser increased the stability of our output loss and FoM progress with training \cite{reddi2018convergence}. A momentum parameter of \texttt{Adam}, $\beta_l$, is used with a value of 0.5 to control the progress of the gradient descent during the training of the network.

        \begin{figure}[h!]
            \centering
            \subfloat[]{{\includegraphics[width=0.5\textwidth,keepaspectratio]{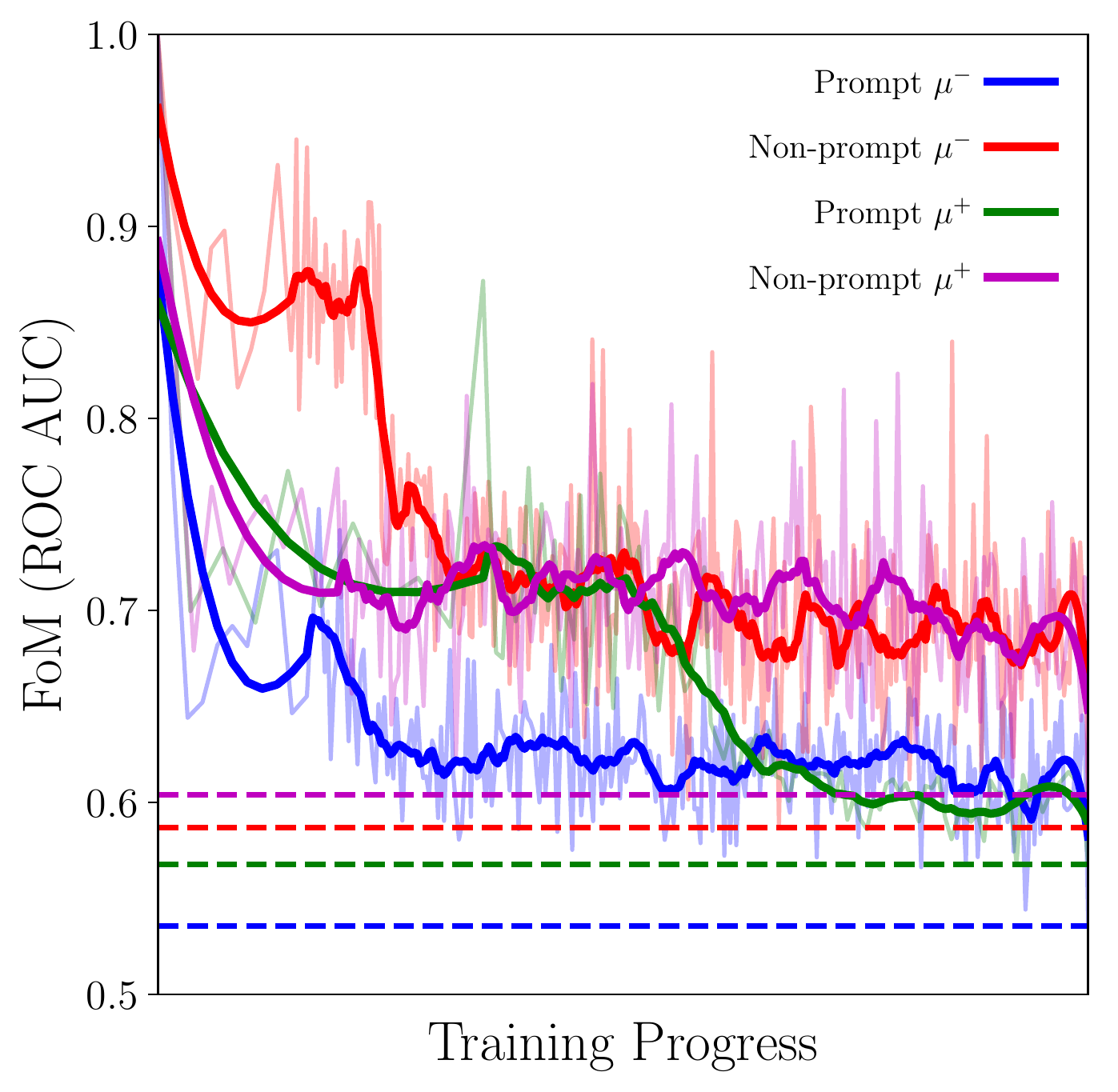} }}%
            \subfloat[]{{\includegraphics[width=0.5\textwidth,keepaspectratio]{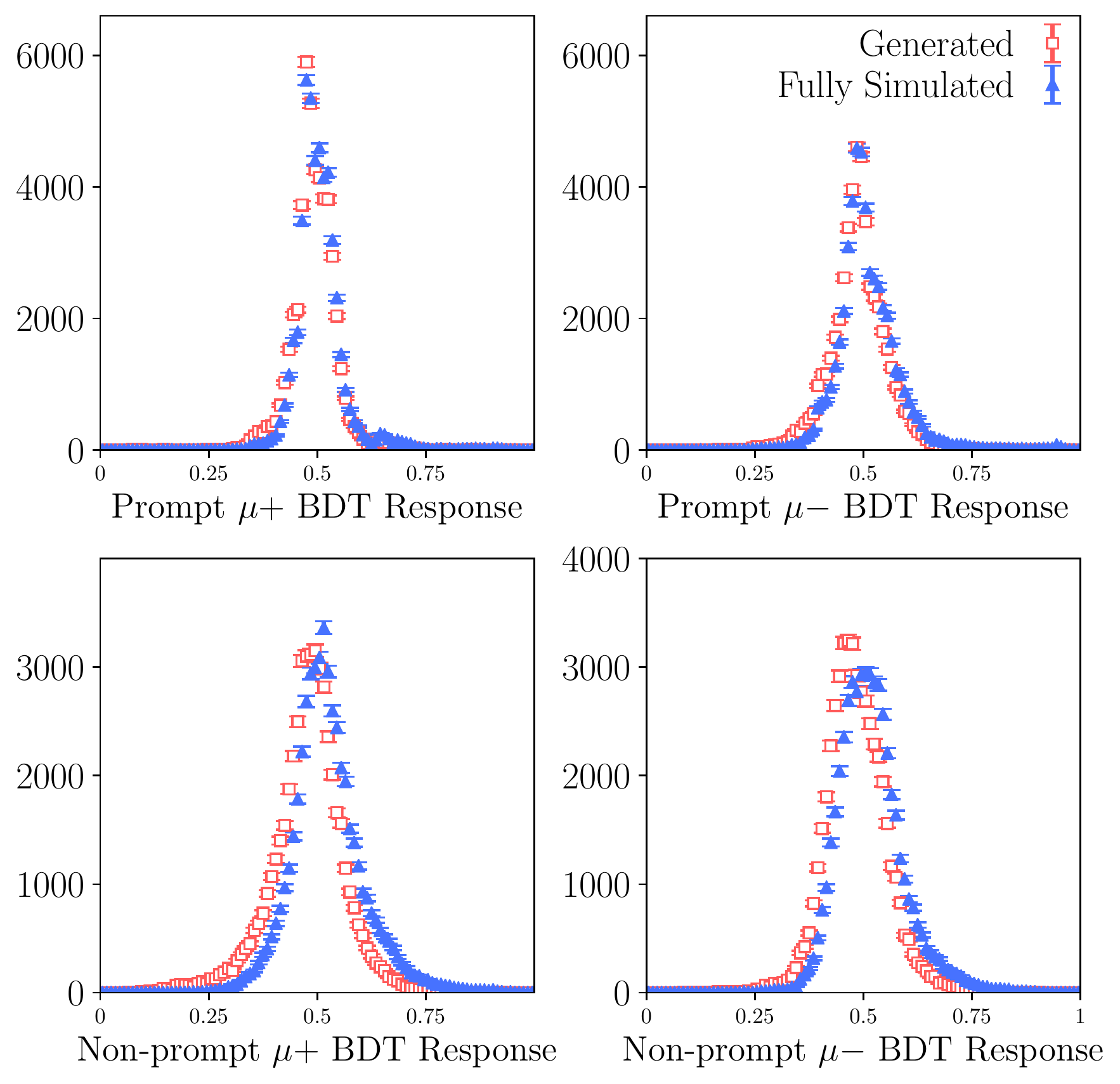} }}%
            \caption{(a) Progress of the FoM ROC AUC value throughout the training of all 4 GANs, raw and smoothed data is displayed. Dashed lines indicate the FoM AUC ROC values of models chosen to generate muons in this paper. Although models were trained past this point this was the lowest FoM AUC ROC value obtained, (b) Distributions of the figure of merit BDT response for both fully simulated and GAN-based muon samples for prompt and non-prompt $\mu^-$ and $\mu^+$.}
            \label{BDT_plot}
        \end{figure}

\section{GAN performance}\label{sec:generatedmuons}

    The progress of the FoM throughout the training of each GAN, as well as the BDT distributions of the optimal GAN models are shown in Fig.~\ref{BDT_plot}. The final FoM values for the prompt $\mu^+$ and $\mu^-$ GAN models are 0.57 and 0.54 respectively. Whereas, the non-prompt $\mu^+$ and $\mu^-$ GAN models return FoM values of 0.60 and 0.59 respectively. A more sophisticated optimisation procedure of the network architecture, such as that suggested in \cite{NN_optimization}, could result in an even better GAN performance.
    
    \begin{figure}[!t]
        \centering
        \begin{overpic}[width=1\textwidth,keepaspectratio]{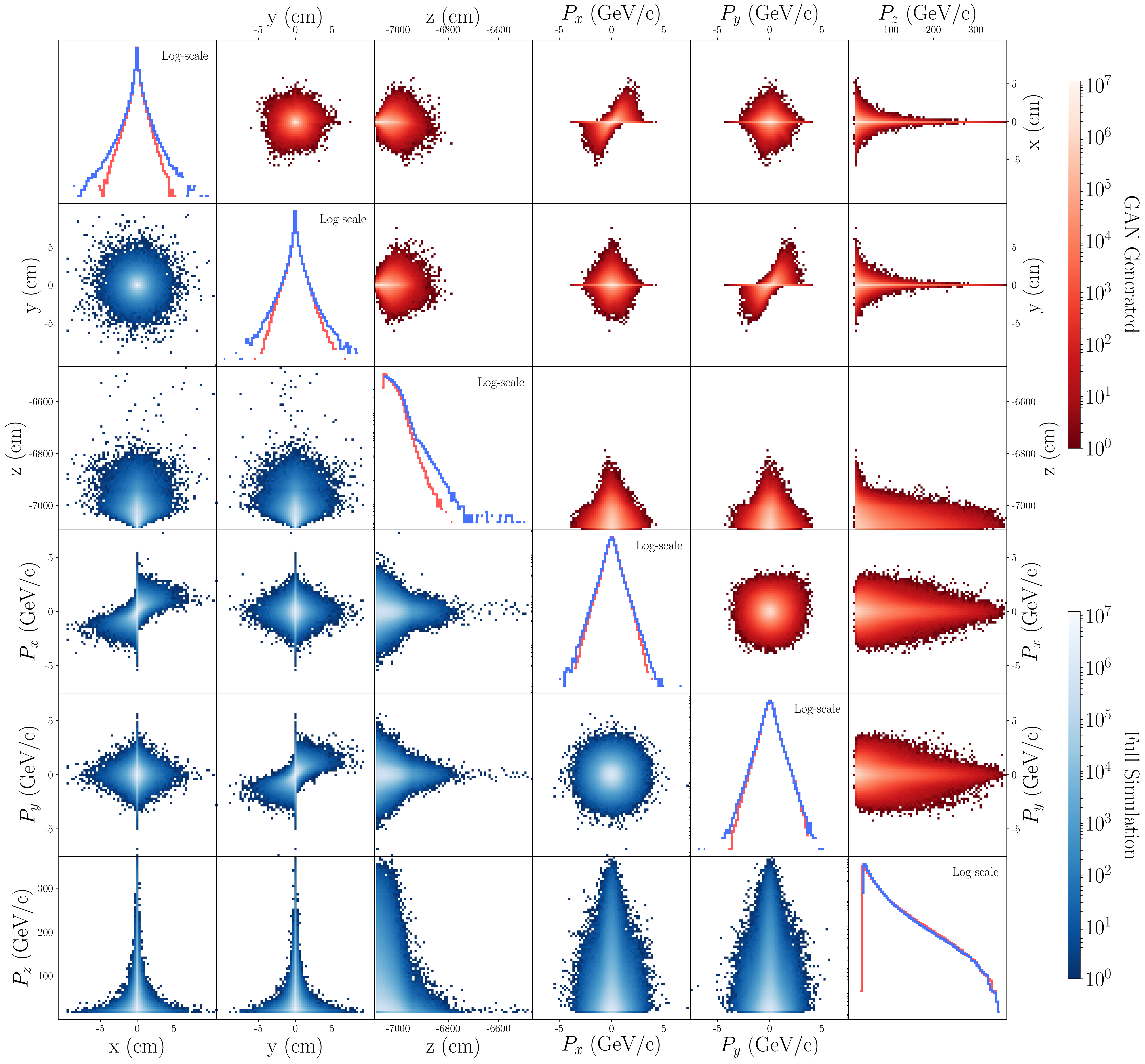}
        \put(5.4,82){\includegraphics[width=0.05\textwidth,keepaspectratio]{SHiP-Full_Black.png}}  
        \end{overpic}
        \caption{Two-dimensional distributions of all unique combinations of muon features for GAN based (upper-half) and fully simulated (lower-half) muons produced in the SHiP target. One-dimensional log scale comparisons of each feature are presented along the diagonal.
        }
        \label{2d}
    \end{figure}

    In order to further visualise the level of agreement between the generated and fully simulated samples, a physical sample of GAN based muons is produced by combining the output from each of the four generators according to the expected production fractions of prompt and non-prompt muons in the simulation. Figure~\ref{2d} compares the one- and two-dimensional distributions of each unique pair of features between fully simulated and generated muons. The GANs can overall reproduce the correct correlations between features, although the tails of the ($x$, $y$, $z$) position distributions are underestimated.

    \begin{figure}[!h]
        \centering
        \begin{overpic}[width=1\textwidth,keepaspectratio]{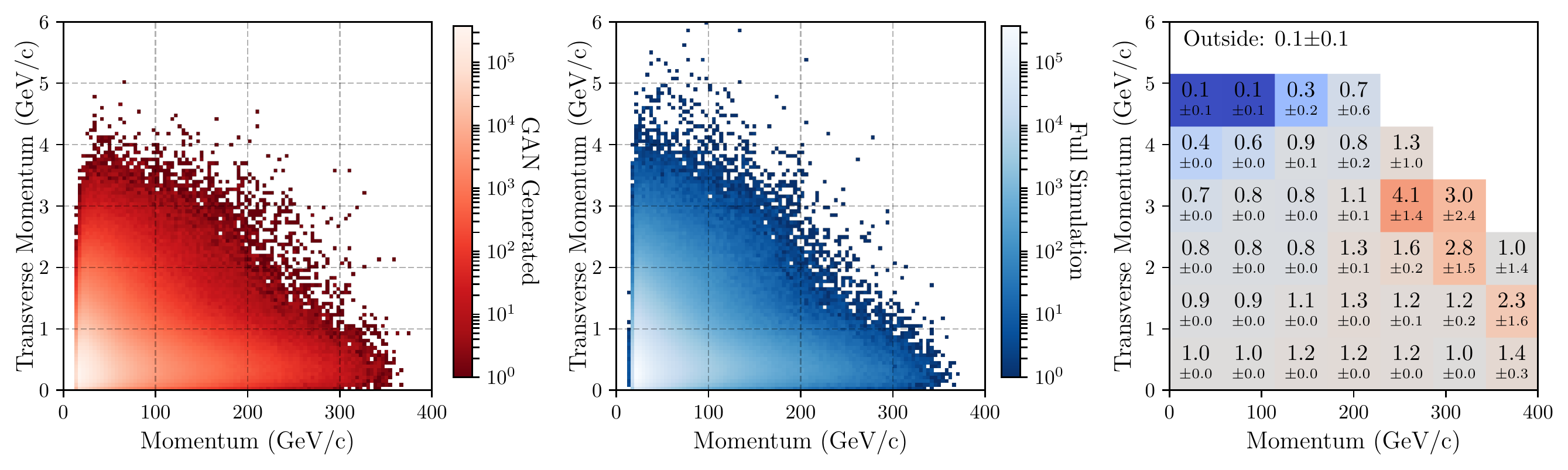}
        \put(57,21.5){\includegraphics[width=0.05\textwidth,keepaspectratio]{SHiP-Full_Black.png}} 
        \put(21.7,21.5){\includegraphics[width=0.05\textwidth,keepaspectratio]{SHiP-Full_Black.png}}
        \put(92.4,21.5){\includegraphics[width=0.05\textwidth,keepaspectratio]{SHiP-Full_Black.png}}
        \end{overpic}
        \begin{overpic}[width=0.55\textwidth,keepaspectratio]{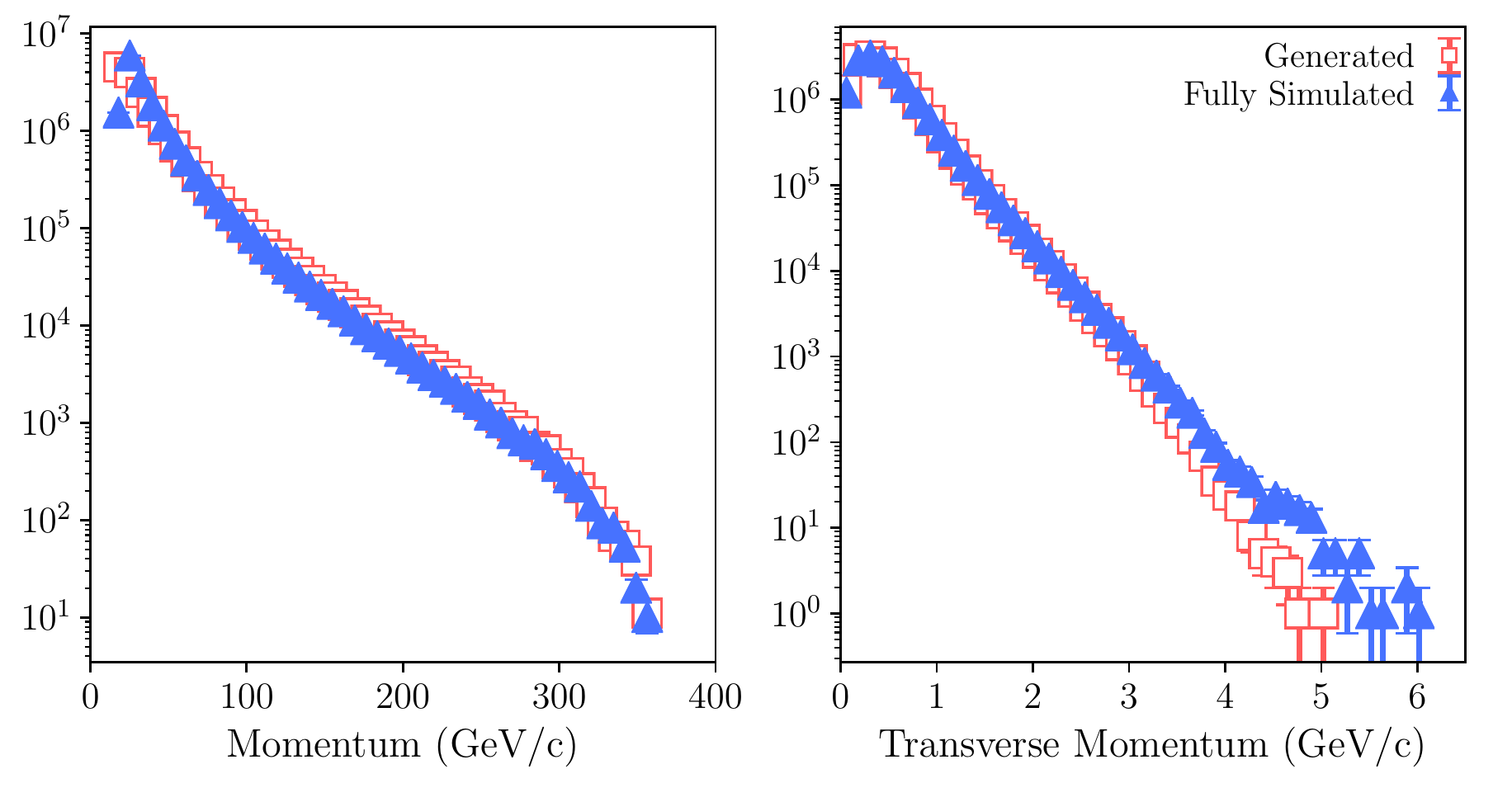}
        \put(8,10){\includegraphics[width=0.05\textwidth,keepaspectratio]{SHiP-Full_Black.png}}  
        \put(58,10){\includegraphics[width=0.05\textwidth,keepaspectratio]{SHiP-Full_Black.png}}  
        \end{overpic}
        \caption{Two-dimensional $p$ vs $p_{\rm T}$ distributions for GAN based (top-left), fully simulated (top-middle) and the ratio (top-right) of muons produced in the SHiP target. The comparisons of the one dimensional projections for $p$ (bottom-left) and $p_{\rm T}$ (bottom-right) are also shown. 
        }%
        \label{mom_plot}
    \end{figure}

    Modelling of the momentum ($p$) and transverse momentum ($p_{\rm T}$) plane accurately is crucial in order to obtain the correct flux of muons reaching the SHiP Decay Spectrometer. Figure~\ref{mom_plot} compares the ($p$, $p_{\rm T}$) plane between the fully simulated and generated samples. The GANs can largely reproduce the correlations between these features, however they particularly underestimate the number of muons with $p_{\rm T}>3$~GeV$/c$. The effect of this mismodelling on the rate and kinematics of muons reaching the Decay Spectrometer is discussed in Sec.~\ref{sec:comp}. To correct the momentum distribution of the generated muons, the three-dimensional ($p_x$,~$p_y$,~$p_z$) distributions of the fully simulated and generated muons are each fit using a three-dimensional Kernel Density Estimator, for example see Ref.~\cite{scott2015multivariate}. For each generated muon, an individual correction weight is derived by taking the ratio between fully simulated over the fully generated KDEs at the corresponding ($p_x$,~$p_y$,~$p_z$) muon coordinate. 
    
    This generative approach accurately models the production of muons in the SHiP target, as long as the kinematic distributions of the muons lie within the phase space covered by the fully simulated training sample. Therefore, samples of muons produced through the GAN are designed to compliment, rather than replace, existing fully simulated samples. By generating vast samples of muons through this generative approach, a better understanding of the performance of the muon shield and the detector response for muons that lie within the kinematic region of the fully simulated sample can be obtained. 
    
\section{Reconstructing GAN generated muons}\label{sec:comp}
    
\begin{figure}[!b]
        \centering
        \begin{overpic}[width=0.85\textwidth,keepaspectratio]{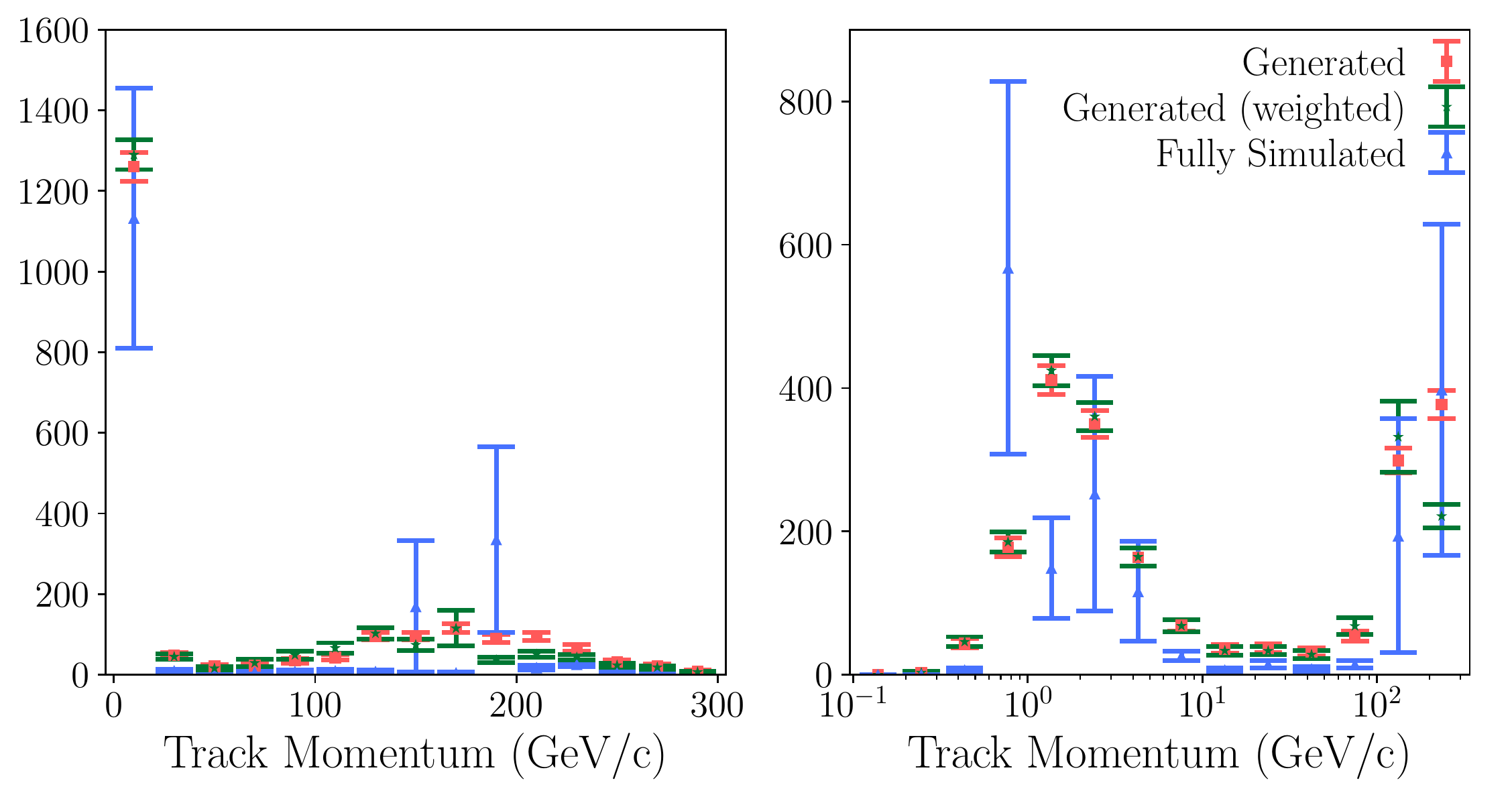}
         \put(58,41.5){\includegraphics[width=0.05\textwidth,keepaspectratio]{SHiP-Full_Black.png}}  
        \put(40,41.5){\includegraphics[width=0.05\textwidth,keepaspectratio]{SHiP-Full_Black.png}}  
        \end{overpic}
        \caption{ Distribution in linear (left) and log-scale (right) of the reconstructed track momentum of muons in the Decay Spectrometer. The distributions of both GAN based and fully simulated muons are also shown together with the effect of the correction to the residual mismodelling of the muon kinematics from the GAN based sample. The distributions are normalised such that they correspond to the same number of protons on target.
        }
        \label{track_mom}
    \end{figure}

    The generated muons are processed using \texttt{FairShip} to simulate their passage through the magnetic shield and the response of the downstream SHiP detector. Figure~\ref{track_mom} shows the $p$ vs $p_{\rm T}$ distribution of reconstructed muon tracks in the Decay Spectrometer of SHiP resulting from the GAN based muon sample. A comparison to the reconstructed muon tracks originating from the full simulation sample is also shown. The effect of the residual correction to the kinematics of the GAN based muon sample discussed in Sec.~\ref{sec:generatedmuons} is found to have a small effect.
    
\begin{figure}[t!]
        \centering
        \begin{overpic}[width=0.85\textwidth,keepaspectratio]{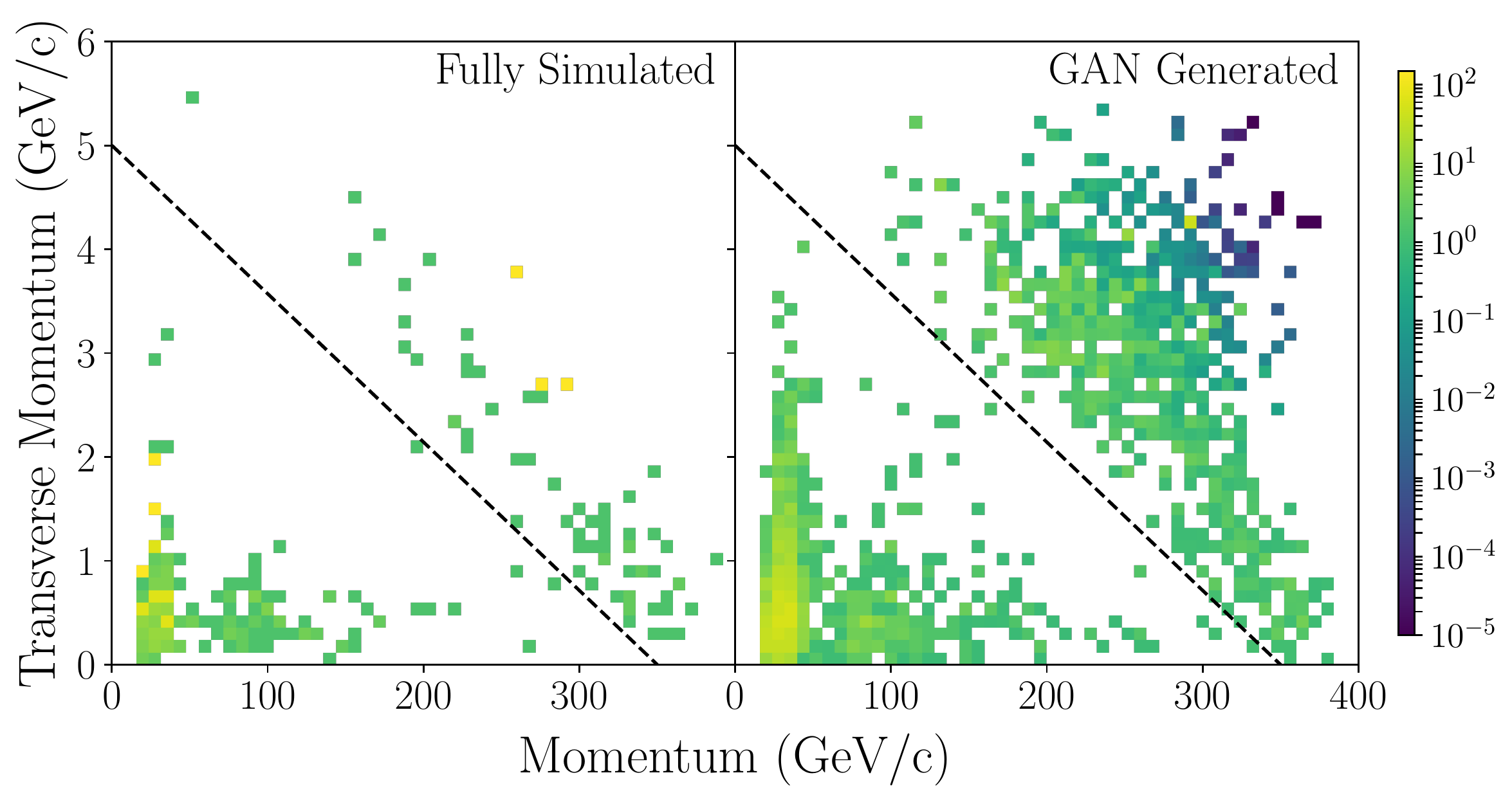}
        \put(14.2,41.5){\includegraphics[width=0.045\textwidth,keepaspectratio]{SHiP-Full_Black.png}} 
        \put(53.2,41.5){\includegraphics[width=0.045\textwidth,keepaspectratio]{SHiP-Full_Black.png}} 
        \end{overpic}
        \caption{Initial momentum of muons passing through the SHiP active muon shield with well reconstructed tracks in the Decay Spectrometer. Full simulation data is presented on the left and generated data on the right. 
        The dashed line indicates the upper region analysed in Table~\ref{rates_table}.}
        \label{track_mom_2d}
    \end{figure}

    Figure~\ref{track_mom_2d} shows the momentum distributions at the production point of the muons, for muons that are reconstructed in the DS. The GAN based and fully simulated muons display similar features in the $p$ vs $p_{\rm T}$ plane. The fully simulated sample exhibits localised hot-spots. These are due to the use of event weights that account for enhancement factors of particular processes that give rise to muons likely to enter the Decay Spectrometer as discussed in Sec.~\ref{sec:ship}. 
    
    The rate of muons that survive the magnetic shield and are reconstructed in the Decay Spectrometer is given in Table~\ref{rates_table}. Both the full rate, and the rate of muons with an initial ($p$, $p_{\rm T}$) distribution corresponding to the upper region of Fig.~\ref{track_mom_2d} agree when comparing the GAN based and fully simulated muon samples. The correction to the kinematic distributions of the GAN based muons discussed in Sec.~\ref{sec:generatedmuons}, changes the rate of generated muons entering the Decay Spectrometer by $\sim4\%$.
    
    \begin{center}
        \begin{table}[!h]
        \centering
        \begin{tabular}{r|p{5cm}p{5cm}}
        Approach & Full Rate (kHz) & Upper Region Rate (kHz) \\ \hline\hline
        Full simulation & $13.9\pm3.4$ & $4.7\pm2.2$ \\ 
        GAN & $15.8\pm0.3$ & $5.5\pm0.2$ \\ 
        GAN (weighted) & $15.2\pm0.5$ & $4.7\pm0.4$ \\ \hline
        \end{tabular}
        \caption{Rates of reconstructed muons in the Decay Spectrometer. The uncertainty on the GAN based muons reflects the statistical uncertainty of the generated muon sample, given the model described in Sec.~\ref{sec:generatedmuons}. \label{rates_table}}
        \end{table}
    \end{center}

 \section{Benchmarking}\label{sec:bench}
 
    With a small expense in the fidelity between the generated and fully simulated sample, the generative approach can produce samples of muons at greater speed. Generating samples of muons from GANs on a GPU provides a speed-up of $\mathcal{O}(10^6)$ relative to the full \texttt{Pythia8} and \texttt{GEANT4} proton-on-target simulation. This test was performed using \texttt{Keras}(v2.1.5) on a \texttt{TensorFlow} backend (v1.8.0) on a single \texttt{Nvidia Pascal P100} GPU card. This speed-up factor includes all the computations required to transform the output features of the generator into physical values. Generating muons using the GAN approach on a CPU is an order of magnitude slower than on a GPU. 
    
    Table~\ref{tab:bench} summarises the results of this performance test. The gain in speed using the generative approach is partly due to the small production cross-section of muons with $p>10$~GeV$/c$, requiring $\mathcal{O}(10^{3})$ proton-on-target interactions to be simulated through \texttt{Pythia8} in order to generate a single muon.

    \begin{center}
        \begin{table}[!h]
        \centering
        \begin{tabular}{p{4.4cm} p{4.4cm} p{4.4cm}}
        Target simulation method & Muons produced in 5 \newline minutes & Time to simulate single muon (s)\\ \hline\hline
        \texttt{Pythia8} and \texttt{GEANT4}& $\sim1$ & \num{1.1e-1}\\ 
        GAN (CPU) & \num{7.5e5} & \num{4.0e-4} \\ 
        GAN (GPU) & \num{3.5e6} & \num{8.6e-5}\\ \hline
        \end{tabular}
        \caption{Summary of benchmarking results. \label{tab:bench}}
        \end{table}
    \end{center}

\section{Conclusion}

    This paper demonstrates the success of using a modern machine learning method to approximate the output of a complex and computationally intensive simulation of muons originating from SPS protons impinging on the target of the SHiP experiment. The GAN models presented in this paper produce samples that emulate the characteristics of the fully simulated sample and can approximate the kinematic correlations of the muons produced in the SHiP target. Furthermore, muons generated by these GANs correctly describe the expected flux and kinematic distributions of muons that survive the magnetic shield and are reconstructed in the Decay Spectrometer of the SHiP detector. 

    The generative models developed in this paper can produce muons $\mathcal{O}(10^{6})$ times faster than the current \texttt{Pythia8} and \texttt{GEANT4} simulation of the SHiP target. However, the muons produced by the generative model are only representative of regions of phase space populated by the full simulation of the target. These generative models are not capable of accurately extending the tails of their training distributions, and are not intended to replace the fully simulated background sample. Generated muons can be used in parallel to complement ongoing background and detector optimisation studies, where this approach can offer a vast increase in the sample size of statistically limited muon background studies at SHiP. Finally, the generative approach presented in this paper can be used to produce muons according to a model trained directly on real data, such as that from the recent muon-flux beam-test campaign of the SHiP collaboration~\cite{ship_test_beam}. Such an approach circumvents the challenge of tuning the multitude of parameters that control the simulation in order to match the data.

\section{Acknowledgements}

    This work was carried out using the computational facilities of the Advanced Computing Research Centre, University of Bristol - \textnormal{\href{http://www.bristol.ac.uk/acrc/}{http://www.bristol.ac.uk/acrc/}}. We would like to thank the NVIDIA corporation for the donation of a Titan Xp which was used for this research.

    \noindent The SHiP Collaboration wishes to thank the Castaldo company 
    (Naples, Italy) for their contribution to the development studies of the 
    decay vessel. The support from the National Research Foundation of Korea 
    with grant numbers of 2018R1A2B2007757, 2018R1D1A3B07050649, 2018R1D1A1B07050701, 2017R1D1A1B03036042, 2017R1A6A3A01075752, 2016R1A2B4012302, and 2016R1A6A3A11930680 is acknowledged. 
    
    \noindent The support from the FCT - Funda\c{c}\~{a}o para a Ciência e a Tecnologia of Portugal with grant number CERN/FIS-PAR/0030/2017 is acknowledged. The support from the Russian Foundation for Basic Research (RFBR), grant 17-02-00607, the support from the TAEK of Turkey, and the support from the UK Science and Technology Facilities Council (STFC), grant ST/P006779/1 are acknowledged.
    
    \noindent We thank M. Al-Turany, F. Uhlig. S. Neubert and A. Gheata their assistance with FairRoot. We acknowledge G. Eulisse and P.A. Munkes for help with Alibuild.
    
    \noindent We thank M. Daniels for his contributions to the construction of the liquid-scintillator testbeam detectors.

    \noindent The muon flux and charm cross section measurements this summer would not have been possible without a significant financial contribution from CERN. In addition, several member institutes made large financial and in-kind contributions to the construction of the target and the spectrometer sub detectors, as well as providing expert manpower for commissioning, data taking and analysis. This help is gratefully acknowledged.

\setboolean{inbibliography}{true}
\bibliographystyle{ship}
\bibliography{main}

\end{document}